# The Information and the Matter – v2++

S.V. Shevchenko,    V.V. Tokarevsky

## Abstract

In this article a revised, to some extent, version of the Information concept as utmost fundamental essence ([1], "The Information and the Matter", v1) is presented – a little more logical grounds and may be the philosophy, a correction of the gravity force concept, etc., as well as some comments to other corresponding articles that were issued in 2007.

### 1. TO THE DEFINITION OF THE CONCEPT OF INFORMATION

There is rather interesting fact that the discussion – "but what is the Information after all?" in the scientific, technical and philosophical literature holds many years already in the various directions without any agreed result.
Without going into the analysis of the discussion in detail (some discussion see [1]), note only, that the discussion's productivity appeared as rather poor, from what follows, e.g., the variety of the definitions of the Information that exists till now in variety of the published works. Though  most of the definitions correspond to "intuitive"  one:

"The information (lat. informatio – an examination, a notion, a concept) –1) a report, a notification about a state of affairs or about something else that is transmitted by a people; 2) decreased, removed uncertainty as a result of the notation obtained; 3) a notation inherently relating to a control, the signals in the unity of its syntactic, semantic and pragmatic parameters; 4) transmission, reflection of the variety of any objects and processes (of alive and non- alive nature)".
Note, that the information appears always only if on a set of some elements at least two logical operations are defined: "identically equal" and "not equal identically". i.e. there is an alternative, or "a bit of information".

(1)  Preliminary reasoning

Let be some set of the elements and on this set a language is defined that include at least the statements (some "information"):  "there is an element";   "there is no elements"; "there can exist some element that doesn't belong to this set ".
All these statements can be "correctly experimentally detected" at considering any ready  set, e.g. – the set of integers from one to two, the set of fingers on a hand, etc. And so "detected"  set doesn't require any verification that it indeed exists, as well as that an information exists relating to the set also.
After choosing another set – the statements   "there is/ no other sets" becomes experimentally also grounded and so become be true.  Considering the sets further one can logically conclude that there is the statement "there is no anything" that relates to the null set.

From  (1) it follows:

(2) - that the corresponding  information exists for any set, including the information for the null set, but in last case the statement "there is no anything" is not true since there is at least the information that there is no anything. True statement is "there is no anything besides the information that there is no anything, besides …"

(3) – from that even in the absence of anything there exists, nevertheless, some information, it follows, that for the information there is no necessity in a "storage device". Indeed, if in a place there are no devices, in this place exists true information "there are no devices for the information". Moreover, in this place there is also true information about any/every (all possible in the place) storage devices and about any/ every data that can be written in these devices.

(4) – since the information can exist without any storage devices, when a device is evidently necessary, then only one situation is possible – when an information is the storage device for an information.
There were for a long time a large number of the attempts to define the concept "Information" (details – see, e.g., [1]) through something more general, but any invented definition eventually turned out to be a tautology – "the information is the information". And negative result of this many-year "experiment" shows that with almost full confidence we [1] can conclude:
(i) – it is necessary to differ the notion "information" on two meanings: the information as some data, and the Information as some concept,
(ii) –an information is the storage device for an information; and
(iii) *the concept of the Information is utmost general and so can not be reduced to anything what is more general. So Information can be defined only through the Information itself, or (but what is evident - incompletely) through the Information's properties.*
Though (ii) and (iii) – that the information can be defined only through itself and that it is the carrier of itself - are, as a matter of fact, the same.

It is evident, that the statements "that is an element of a set" and "there is no this element in this set" are practically identical relating to full (may be – infinite) information about this element – the difference is only in the indication of the belonging of the element to given set.. From this it follows that the cyclic statement "there is no anything besides the information that there is no anything besides…" (further "Zero statement") is the logical singularity and it contains all/full information about any/everything.

(5) From (1) – (4) it follows, that **conceptually the Information is some specific infinite set**. Each element, including Zero element, of this set is a bit "I/not-I", and so any element contains all others, so the Information in some sense formally may contain from one element; but since "not-I" part contains infinite number of elements this approach seems not too useful. Each element contains all/full information about it's "life" in the "past", "present" and "future" as well as all analogues data about every/ all elements in the set Information. Any element in "here and now" state contains all it's states in the part "not-I", though specifically, of course.

(6) Since the Information elements can exist only as a number of logical connections and realizes as a choice of some alternatives, the Information set must be "countable" set, but the set's cardinality is utmost maximal. It is well known that the cardinalities of the "countable" set "(2^N)" at $N \to \infty$ and the cardinality of the continuum are equivalent. It is rather possible (but that may be rather interesting task to prove) that the Information set is a set "2^(2^(2^….(2^N)…)" where both – $N$ and the number of parenthesises are infinite.

(8) It is known that on the set "Information" can exist some - at least finite - subsets, where the subset's elements are connected by logical conditions and, so, on the subset there exist some symbolic language. From that it follows - since a language exists on the Zero element and since the Zero element coincides with the (full) Information set - than the contain of any subset of the Information can be expressed by using some language.

Though, of course, - with Gedel's incompleteness constraint – if the rule (operation) set and alphabet in this language are finite, then there doesn't exist a full system of the concepts – "in any language one can always find a true statement that can not be proved in this language".

Note, that Gedel's constraint is valid to any finite Information subset (further – the informational system, IS), from what follows, that there are no finite subsets in the reality**.**

It may be necessary to add, that the Information set contains also any "possible" false information.

## 2. INFORMATION AND THE PROBLEM OF CREATION/ EXISTENCE OF THE MATTER

There are two options for the existent material World (Universe):
(i) – the World exists infinitely, without the Beginning and the End in the Time, and
(ii) - the World exists in some finite time (and, what is not impossible, without the End in the Time)

The choice between the options – *is a question of belief*, because now there is no data that could be sufficient evidence of what option is true. However, if the option (ii) is true, then it follows that until the Beginning:
(a) - there existed true information in the form of recurring infinite (*but consisted of finite set of concepts or a language's symbols*) "Until Beginning" statement (UBS) "there is no this World besides the information about this World", or, if our World is unique, - **"there is no anything besides the information that…"**;
(b) - nothing else, besides the information, existed;
(c) – As it was pointed out earlier, the information in the UBS was infinite and contained all information about all, including the information related to the creation and evolution (including all version of the evolution in future) at least of our World.

For the option (ii), from (a), (b), and (c) strictly follows, that:
(I) – our material World was generated from the Information; and
(II) – since the World formed from the Information then now it is a system that consists of the (including - ordered) informational structures (IS).

Note, also, that for the option (i), the statement, that the World is a system that consists of the IS, is not prohibited.
I.e. the World has a similarity to a computer. The human does not observe these IS directly and does not read "initial information" till now (in a similar manner a human does not observe the switching of the logical elements in the chips of a computer and sees only the pictures on a monitor). He sees (registers by the instruments) a result of the work of a "software" that was developed by the Nature. The main objective (and it is not impossible – the result) of proposed here "informational" conception is the further detailed elaboration of our ideas about the World.

The Information in the World develops (comes to be) as a complication of the physical systems at the interactions of the simplest IS – the elementary particles. Further assume, that these IS consist of some fundamental logical elements (FLE) of the information – an analogues of the logical elements in a computer, when in these FLEs some "primordial" logical statements are used instead of transistors, diodes, etc**.**
The hypothesis that the World is some computer is not new – the assumptions that the World is a Computer appeared practically at with the same time as the widening of the computers application, i.e. in 50 – 60 of XX century [2], [3]. However the main problem – what is "*the Ultimate* (Universe) *Computer*" (E. Fredkin, [3]); how it could be originated; what could be its "elemental base"; how it could be controlled, etc., - were not solved till now. For example in [3] the statement: "As to where the *Ultimate Computer* is, we can give an equally precise answer; it is not in the Universe - it is in an *other place* " - is one of the principal conceptions of the E. Fredkin "digital philosophy". A non-productivity of such a conception is obvious, however, in these works there was considered the critical for the *Ultimate Computer* existence *problem* – the problem of the creation and work of the "inertial" computers**,** i.e. *the computers that do not dissipate (so do not consume) the energy, and it was shown that such a computers can be indeed made*. It can be noted, also, that "World Computer" differs from, e.g., PC - seems that it resembles some alive constitution where different ISs, including the fundamental elementary particles, resemble the organic cells and exist to some extent independently, not under control of a "central processor"

### 3. SOME CONSEQUENCES FROM THIS INFORMATION CONCEPT – A MODEL

*3.1. Big-Bang energy*

One of the main problems of physics is *the problem of the Universe's creation* becomes much clearer. Now there is a lot of evidence that may be explained at most logically if the hypothesis that the World was originated as a result of a "Big Bang" in a point having "energy singularity". The main problem of this hypothesis is the deficiency of starting energy $10^{85} - 10^{90}$ MeV – is taken away in the informational concept, since the logical singularity of the UBS is infinite and the true information in the UBS was enough to create the World at the "Big Logical Bang".

*3.2. Particles identity*

It makes understandable one of the fundamental postulates of the quantum mechanics – the postulate that all elementary particles of given kind are identical when this postulate is definitely in contradiction with other fundamental QM postulate – about the principal stochasticity of the processes in the micro world. *Now there is only one case known where it is possible to ensure the identity of something, viz. - the identity of the information*. And any informational structure can be reproduced in any number of copies. So it is possible that the elementary particles are just some informational clones. At that the wave function is some mathematical representation of the future information - that always exists - about/ for a particle in the spacetime.

*3.3 Language and reality*

It becomes understandable the startling adequacy of the languages of the mathematics, physics, other sciences for the description of the processes in the material world.

3.4. *The Space and the Time.*

It follows from the experience that the information can exist in two kinds – a specified (fixed) information (e.g. – ascertaining/ description of some fact; a computer code listing, etc.) and a dynamic information (e.g. –data processing on a computer; changes of state in a physical system). from the fact that the Information is countable it follows that Space, Time, Matter, Energy and all rest in the World must be something discrete
The different informational structures in the World should consist of the different FLEs, because of different ISs are informatively separated. *The informational separation appears for an observer as the "space" separation. The populations of the structures FLEs (further – "t-FLEs") and of the "ether" FLEs (further – "e-FLEs") as a whole is the Space. It is evident, that the Space is discrete (quantified).* This hypothesis isn't new – some analogue was the presentation of the Space as a "spin network" (in this concept – as a "FLE network") by R. Penrose [4].

The fact that the space-time continuum hypothesis contains some insuperable logical self-contradictions was cleared up already 2500 years ago when Zenon stated his aporias. Now to the same conclusion also the physics goes when it faced with various divergences at the development of the field theories.
Because the interactions between elementary ISs are random and small as well as since the FLE's dimensions are very small too, the Space and the Time are observed (till now) as they are uniform and continuos.

### *3.5. The "development" (the realisation) of the Information in material World*

So the informational approach means that anything in the World is transformations and interactions of the ISs and that the elementary particles are some primary ISs also. Correspondingly, and taking into account the "wave nature" of the particles in the Space and in the Time which follows from Schredinger equation, we can built two options of the informational currents (IC) –the time IC and the space IC and one option for fixed information using only the most common physical parameters and Dirac's (**note, that in Refs. [1], [5] –[7] -Planck's constant is used**) constant, $h_D$:

- the time IC (t-IC):

$$j_t = \frac{1}{h_D} \gamma m_0 c^2, \quad (1)$$

- the space IC (s-IC):

$$j_x = \frac{1}{h_D} \gamma m_0 c^2 \beta^2, \quad (2)$$

- fixed information:

$$\Delta I_M = \frac{\Delta M}{h_D}. \quad (3)$$

($c$ is the speed of light in vacuum; $\beta = v/c, \gamma = 1/(1-\beta^2)^{1/2}$ is the Lorentz – factor of a particle movement, $\Delta M$ –is angular moment, $m_0$ is the rest mass. The dimensionality of the time and the space currents – [bit/s], the dimensionality of fixed information –[bit]).

### *3.6. The elementary particles as the informational structures of the material World*

As it is assumed now in the physics theory (there are some experimental evidences) that the elementary particles are divided into fundamental and derived (that are composed of the fundamental particles). In turn the fundamental particles include the quarks and the leptons. Besides there is so called "exchange" particles (mediators), i.e. the particles that are necessary to mediate some interactions – the photon, the gluon, the mesons (at the strong forces).

In the proposed model all particles (as any other material structure) are some informational currents and its values are determined by Eqs. (1)-(3).
It is well known that the transformation of the information in physical (e.g., - a computer) and biological systems needs rather essential energy expenses. As it was noted here above, in the works of E. Fredkin, T. Toffoli., N. H. Margolus (see, e.g. [3]) as early as 70-80 years of former century it was shown that it is possible to make "inertial" computer which works without the dissipation of energy and some versions of logical elements were suggested to built such a computer. One *of the utmost important condition* when the computer could be made *is the reversibility of its logical elements*. I.e. this computer should have the possibility to work in both – in a direct and in corresponding reverse order**.**
Turning back to the problem of the t-ICs of the (at least of stable) elementary particles it is reasonable to suppose that the particles are some cyclic informational structures with rather short (since the particles periods experimentally are not observed till now, *except the one for the photon*) periods of the $2\pi$ t-IC – steps, when the structure's logical elements (t-FLEs) are reversible. So if the particles t-ICs are some direct algorithmic codes then corresponding antiparticles are the codes with reciprocal code order. At that the antiparticle "lives" as if in the negative time as it was assumed by P. Dirac already in the first half of past century. Experimental discovery of the antiparticles practically for any particles, with the exception of some mediators (but it is not fully correct since the mediators are, as it seems, some integrated "particle + antiparticle" combinations – see [1]) indicates that the conjecture that the Nature's logical elements are some analogous of Toffoli – Fredkin ones seems quite reasonable.

### *3.7. Elementary particles and mediation of the forces*

Now four kinds of the interactions (4 forces) are known – gravitational, weak, electromagnetic (EM), strong which differ, e.g. for the proton as (approximately) $10^{-36}:10^{-3}:1:10^{3}$.
In suggested model the mediating particle appears just naturally as this particle is some intermediate algorithm aimed to synchronize and to unify in some part (on some time) the steps of t-ICs of interactive particles, if it is possible. *Then the rate of unified steps of particles is the "binding" (really – potential energy) energy of the system, when at each unified step interactive particles get the momentum $p_0 = \pm h_D / r$* [1]. The "binding" energy (unified t-IC) here is not "usual" binding energy in physics or chemistry, but these energies become equal, sometimes in the case of non- elastic interactions, e.g. – after two bodies "sticked together" under gravity and its temperature returned to the one before the interaction. If there is a possibility for a body to move, the unified t-IC will be the same, but "usual" binding energy will be less on the body's kinetic energy.

### *3.8. The gravity and electricity*

There are a number of theoretical gravity concepts now that appeared since famous Newton work. The main drawback of the Newton theory is that it is true in a "static" (or non-relativistic) case only. When a body becomes to move, Newton's law becomes invalid. To overcome the problem of gravity forces at a movement there were developed a number of theories, which are based on the analogy of Coulomb and Newton laws and, further, on the corresponding analogy with Maxwell's theory of EM force. The first attempt, probably, was the O. Heaviside work [8], another – and utmost known – is the General Relativity theory.
All these works contain the main – and, as it seems, unresolved till now – problem "What is analogue of E-M magnetic force?" In GR so called "gravitomagnetic" force is introduced, but the analogy with the EM exists only for week field approximation. Such a situation leads to some strange conjectures as, e.g., the "frame dragging"; there are some articles where it is shown, that GR concept leads to conjecture that there exists some "critical" body's speed when gravity force changes the sign: "…This equation contains a critical speed $v_c = 1/(3)^{1/2}$; that is, for motion with $v < v_c$, we have the standard attractive force of gravity familiar from Newtonian physics, while for $v = v_c$, the particle experiences no force and for $v > v_c$ the gravitational attraction turns to repulsion. These results are valid in the linear approximation for the gravitational field…" [9]. Another example – the work [11], where so called "Gyron field" conjectured.
To test a gravity theory it is reasonable to consider, first of all, well known standard problems (e.g.[10]), one of its is the problem of the forces between two moving bodies having charges and masses, as, e.g., that was made in [11], [12].

### *3.8.1. The gravity and EM: static solution*

#### *3.8.1.1 The gravity*

Remaining in the informational concept it is possible to put forward [1], rather reasonable conjecture that it seems - since the gravity force is universal (regardless to the kind of particles) and acts always as an attraction - there is only one case to satisfy to this condition: if the gravitational binding energy of a system of some bodies *is proportional to the accidental coincidence rate of the t-ICs of the* *particles* of these bodies *that always exists if the t-FLE's "switchings" time is not equal zero*. From this suggestion it follows, that: (i) - the gravity force should be very weak, and (ii) – the gravitational force between some bodies is a result of the interactions of smallest (that aren't divided on a components in real conditions) particles that constitute the bodies.

For two bodies having masses $m_1$, $m_2$, placed on the distance between the bodies, r, the "Newtonian" binding energy is

$$E_{gN} = G\frac{m_1 m_2}{r}, \qquad (4)$$

where $G$ is Newtonian constant of gravitation.

Basing on assumptions above, it can be shown that, under rather plausible assumptions, the binding/ potential gravity energy can be expressed as follows.

Let these bodies consist of some numbers of different elementary particles, e.g. body 1 consists of the $I$ kinds of particles and body 2 – of $K$ kinds of particles having, correspondingly, the masses $m_{10i}$ и $m_{20k}$. So the bodies masses are equal

$$m_1 = \sum_{i=1}^{I} N_i m_{10i} = \sum_{i=1}^{I} m_{1i}, \qquad (5)$$

$$m_2 = \sum_{i=1}^{K} M_i m_{20i} = \sum_{i=1}^{K} m_{2i}. \qquad (5a)$$

Assume further that the t- (at least start/stop) and e-FLE's sizes are equal to Planck's length, $l_P$.

Assume, also, that at every t-IC step in the Space a "rim" of e-FLE's "switchings" starts expand with radial speed equal to the speed of light, $c$, so the rim's area is equal $2\pi r l_P$; and
- the times, $\tau_t$, of the t-FLE's "switching" and of the interaction of the e-FLEs and t-FLEs, $\tau_r$ are the same and are equal $\tau_i = \tau_r \equiv \tau = l_P/c$.

Then the accidental coincidence rate in the body 2 when radiates the informational current of the body 1, $N_{cc21}$, is equal:

$$N_{cc21} = \sum_{i,k} N_i M_k \frac{m_{10i} c^2 \cdot 2\pi r l_P}{4\pi r^2 h_D} P \frac{m_{20k} c^2}{h_D} 2\tau; \qquad (6)$$

where $P$ – is the probability of particles' t-FLEs interaction if a rim of the particle $m_{20i}$ t-IC FLE switching passes through the t-FLE of the particle $m_{10k}$.

Since the system is symmetrical, the sum coincidence rate of both bodies is equal $2 N_{cc21}$ and the binding (potential) energy is equal, if the probability $P$ is equal 1/2:

$$E_{gI} = 2h_D \cdot N_{cc21} = \frac{c^3}{h_D r} l_P^2 \sum_{i,k} N_i M_k m_{20i} m_{10k}$$

$$= \frac{c^3}{h_D r} l_P^2 m_1 m_2 \qquad (7)$$

$$E_{gI} = 2\pi c p_0 j_{1t} j_{2t} \tau_P^2 = 2\pi c p_0 <\Delta I_1><\Delta I_2>$$

where $p_0 \equiv h/r$; $<\Delta I_{1,2}>$ – are the mean increments of the bodies t-ICs for one Planck time.

From Eq.(7) it follows that $E_{gI} = E_{gN}$.

It is evident that from Eq.(6) follows equation for the gravity force in statics:

$$f_{G0} = \frac{dP}{dt} = 2N_{cc21}p_0 = \frac{Gm_1m_2}{r^2} \qquad (7a)$$

From Eq.(7) it follows also, that the value of the gravitational (coincidence rate) current in any particle is random in the time. Correspondingly so does the gravitational force that impacts on this particle and some uncertainty should appear at gravitational interaction of small masses. The detection of corresponding randomness of the gravity force (or some equivalent physical value) will be rather weighty evidence that suggested informational concept is true (possible experiments – see [1] and Refs. in [1]).

It can be noted here, that from Eq.(7) follows, as it seems, some interesting result.

If there is a couple of bodies having masses $M$ and $m$, $M>>m$, then the fraction, $p$, of the gravitational coincidence t-IC in $m$ is equal

$$p \approx \frac{j_G}{j_m} = (\frac{GMm}{rh_D})/(\frac{mc^2}{h_D}) = \frac{GM}{rc^2} \qquad (8)$$

It is rather reasonable to conjecture, that when in an IS, e.g., in an atom, a "gravitational coincidence" happens, then corresponding IS's t-IC step is not used in the IS's algorithm. So for this IS it's "own" t-IC becomes slowed down and the IS becomes "to live in slowed time", $t_{own}$ that is evidently inversely proportional t-IC and is equal in first approximation to:

$$t_{own} \approx t_0(1 - \frac{GM}{rc^2}) \qquad (9)$$

where $t_0$ is the IS's time for free body. This value is, in turn, approximately equal to the gravitationally dilated time in GR:

$$t_{GR} \approx t_0(1 - \frac{2GM}{rc^2})1/2 \approx t_0(1 - \frac{GM}{rc^2}) \qquad (10)$$

*3.8.1.2 The electricity*

The electric force is rather similar to gravity - both potentials are as $1/r$, if some bodies interact then in reality when the interactions of separated particles occur, etc.; except, of course, that gravity is much weaker that electric and that electric force can act as attraction and as repulsion. So it is rather reasonable to conjecture that the equations for the potentials should be similar also, but the probability of electric interaction should be larger. So for the electricity potential we can obtain an analogue to Eqs. (6), (7) (for a couple of particles with the charge $e$) the equation:

$$N_{cc21} = \frac{m_1c^2 \cdot 2\pi r W_{1E}}{4\pi r^2 h_D} P_E \frac{m_2c^2}{h_D} 2\tau_E, \qquad (11)$$

where $W_E$ – is the "electric rim" width, $P_E$ – the probability of the interaction if through 2-particle a rim of 1-particle passed, $\tau_E$ – the "passing'" time. Under rather plausible conjectures that: $W_{1E} = \sqrt{\alpha}\lambda_{1C}$, where $\lambda_C = h_D/mc$ is the Compton length of a particle; $\tau_E = W_{2E}/c$; $P_E = 1/2$; $\alpha$ - the fine structure constant, we obtain from Eq. (11) that electric potential energy is

$$E_E = 2h_D \cdot N_{cc21} = \frac{\alpha h_D c}{r} = \frac{e^2}{4\pi\varepsilon_0 r} . \qquad (12)$$

and for EM force in statics obtain

$$f_{E0} = \frac{dP}{dt} = 2N_{cc21}p_0 = \frac{e^2}{4\pi\varepsilon_0 r^2}$$
$$= \frac{q_1 q_2}{4\pi\varepsilon_0 r^2} \qquad (12a)$$

(the last in Eq.(12a) is for arbitrary charges).

### 3.8.2 The forces at the movement

Some analogy between the forces keeps also at uniform and straight movement. Indeed, if one choose for a couple of charged masses the magnitudes $m_1, m_2, q_1, q_2$, so that the 2-bodies system is in a balance at static conditions, then at any $r$ and in any inertial frame that moves relating to the "rest" frame with any given speed, $v$, and at any angle between $\vec{v}$ and $\vec{r}$.

So let further consider a system that consists of two bodies (particles) having any charges $q_1$ and $q_2$, as well as any masses $m_1$ and $m_2$; the bodies are placed on a distance $r$. For simplicity consider 2 cases when the system can move with the speed $v$ along the axis $X$: (i) - both bodies have equal $X$- coordinate, $x$; the force is perpendicular to the movement direction, and (ii) - both bodies have zero $X,Y$ coordinates, the force is parallel to the movement direction. The equations for 4-forces for rest frame are Eq.(7a), (12 a):

- the force for the case (i) is, e.g., $(0,0,f_0, 0)$;
- for the case (ii) it is $(0, f_0, 0, 0)$.

The 4-forces at a bodies system's movement we obtain instantly by using 4-vector Lorentz transformations:

- the force for the case (i) is $(0,0,f_0, 0)$;
- for the case (ii) it is $(0,\gamma f_0, 0, 0)$. I.e. – for the case (i) the 3-force$_{rest}$ and 4-force$_{move}$ are equal; when in case (ii) both 3-forces are equal. 3- forces in the frame K when the system $K'$ is the system where the bodies are in rest, will be:

$$f_{E,Gi} = \frac{1}{\gamma} f_{E,G0} \qquad (13)$$

$$f_{E,Gii} = f_{E,G0} \qquad (13a)$$

### 3.8.2.1. The electricity

It is well known, that moving charge creates electric and magnetic fields $E$ and $B$, when

$$\vec{B} = \vec{\nabla} \times \vec{A} = \frac{\vec{v}}{c^2} \times \vec{E} \qquad (14)$$

where $A$ – is the vector potential of the electromagnetic field,

$$\vec{A} = \frac{\vec{v}}{c^2} \varphi \qquad (15)$$

and $\varphi$ is electric potential;

$$\phi(\vec{r},t) = \gamma \frac{q_1}{4\pi\varepsilon_0 r^m(t)}, \qquad (15a)$$

where $r^m \equiv [\frac{(x-vt)^2}{1-\beta^2} + y^2 + z^2]^{1/2}$, in this task problem $r^m = r$;

$\vec{\nabla}$ is del operator; and (perpendicular to axis X at point x) field $E$ is equal:

$$\vec{E} = -\vec{\nabla}\varphi - \frac{\partial \vec{A}}{\partial t} = -\vec{\nabla}\varphi \qquad (16)$$

An observer in the frame where the charges move can measure both – the electric and magnetic fields, at that the electric 4- force which acts on a charge will be equal

$$F_{Ei} = \gamma q_2 E = \gamma^2 \frac{q_1 q_2}{4\pi\varepsilon_0} \frac{1}{r^2} \qquad (17)$$

i.e. the electric (repulsive in this case) force at relativistic speeds rises essentially comparing with the static case ($\gamma=1$). But the 4- forces must be equal independently of a frame and that is made by magnetic force – parallel currents attract (second term below):

$$\vec{F}_{LEi} = \gamma q_2 (\vec{E} + \frac{1}{c^2} \vec{v} \times \vec{v} \times \vec{E}) \qquad (18)$$

and so 4-force

$$F_{LEi} = \frac{q_1 q_2}{4\pi\varepsilon_0} \frac{1}{r^2} = F_{rest} \qquad (19)$$

i.e. – as it is for static conditions.

For the case (ii) the magnetic force is equal zero, but the electric force depends on the vector potential, so this force is equal

$$f_{Eii} = q_2 E = \frac{q_1 q_2}{4\pi\varepsilon_0} \frac{(1-\beta^2)}{r_m^2} \qquad (20)$$

where $r_m^2 = (x_1 - x_2)^2 = r^2(1-\beta^2)$, and

$$f_{Eii} = q_2 E = \frac{q_1 q_2}{4\pi\varepsilon_0 r^2} \qquad (20a)$$

- as it should be in static condition.

*3.8.2.2 The gravity*

Since the gravity end electricity are some resembles, the gravity should have the vector potential also. It is rather evident, that this potential (and corresponding "gravitomagnetic" force) should be caused by the movement and, taking into account that a movement is always accompanied with space informational current Eq.(2) which should propagate through space analogously to the t-IC, we can conclude, that **this potential should be proportional to both – to the momentum and to t-IC**. Note, however, that there is important difference between electric and gravity forces. From Eq.(15) follows that the electric potential (so – and *E* and *B* fields) depends on the movement only "geometrically", i.e. – through the relativistic (Lorentz) transformations of reference frames, because of the dependence on relativistic t-IC increase is eliminated as a result of that $W_{1E}$, $W_{2E}$ are inversely proportional to Lorentz factor also. This fact is used in standard theories as the statement that "electric charge is relativistic invariant" (in fact, the relativistic invariant is square root of alpha). But that is non-correct for a moving mass – it is proportional to the Lorentz factor.

Note, also, that the momentum at a movement plays two roles: at first it relates to the movement, creating, or, more correct – relating to the space current Eq. (2), at second – it "carries" the gravity scalar potential. The last is modified at a movement also – it is reasonable to conjecture that when in static case this potential is proportional to the full time current, at a movement a part of the t-IC "is spending" on the movement, and only the difference of the currents, $j_t - j_x = j_t(1-\beta^2)$ is used for the radiation of "gravity rims". At that the t-IC of other – "receiving"- body is capable to interact with "rims" without losses. So for gravity field the scalar potential is equal:

$$\phi_G(\vec{r},t) = \frac{G(\gamma m_1)(1-\beta^2)}{r^m(t)} \quad (21)$$

and here, as for the electricity potential above, $r^m = r$. The gravity vector potential is

$$\vec{A}_G(\vec{r},t) = \frac{\vec{v}}{c^2}\varphi_G \quad (22)$$

Then the "gravitoelectric" (perpendicular to axis X at point x) field in the frame where the charges move is equal

$$\vec{E}_{Gi} = -\vec{\nabla}\varphi = \gamma^2 \frac{Gm_1(1-\beta^2)}{r^2} \quad (23)$$

and the "gravitomagnetic" field is equal

$$\vec{B}_{Gi} = \vec{\nabla} \times \vec{A}_G = \frac{\vec{v} \times \vec{E}_G}{c^2} \quad (24)$$

Executing further the calculations analogously to Eqs. (16) –(20), we obtain that gravitoelectric 4-force is equal

$$F_{GEi} = \gamma(\gamma m_2)E_G = \gamma^2 \frac{m_2 G \gamma^2 m_1}{r^2}(1-\beta^2) \;, \quad (25)$$

what is evidently larger then the gravity force in static case. But parallel mass currents (always – in contrary to [9]) repulse; and gravito – Lorentz 4-force is equal

$$\vec{F}_{Gi} = \gamma^2 m_2(\vec{E}_G + \frac{1}{c^2}\vec{v} \times \vec{v} \times \vec{E}_G) = \frac{Gm_1 m_2}{r^2} = F_{Grest}. \quad (26)$$

- as it should be in the static conditions situation.
For the case (ii) obtain analogously to the electric force:

$$f_{Gii} = \gamma m_2 E_G = \gamma^2 \frac{Gm_1 m_2}{r_m^2} (1-\beta^2)^2 \qquad (27)$$

$$f_{Gii} = \frac{Gm_1 m_2}{r^2} \qquad (27a)$$

- as it should be in static condition.

The similar inferences for the gravity force was obtained in [11], [12] at the consideration analogous problem of the equality of electric and gravity forces between two bodies at the rest and at the movement. In [11] the factor $(1-\beta^2)$ was introduced as the appearance of an additional "Gyron force"; in [12] – the result is explained as a sequence of that the rest mass, as the charge, is the relativistic invariant also. The last is evidently non-correct – simply in this test problem such a conjecture is valid by chance. As to some additional forces, then from the consideration above follows, that, it seems, there is not a necessity to introduce its in the theory of gravity.

* * *

So it seems that the particles' algorithms are, as a rule, some closed-loop sequences that contain some t-FLEs which acts with Space e-FLEs and become sensitive to external e-FLEs $2\pi$ times for a cycle, at that – for the gravity such logical elements are single when for another forces the algorithms contain some "sensitive" "charge algorithm's" sections. For electricity such a section is $\sim\alpha^{1/2}$ of full algorithm's length. It seems, also, that in the full algorithm's length for a particle in the rest only small part is "informational" – it is "diluted" by a "ballast" t-ELEs. If the particle starts to move, the t-IC rises due to that "ballast" t-ELEs drop out the sequence. Some algorithms aren't "closed", e.g. photon's one is rather possibly "linear". To make the cycle in this case it is necessary to make some "there and back" or "caterpillar" sequence what indicates on that the photon (as well as, it seems, any other mediator) should be a composition of two fermions [1].

Some other physical corollaries can be found in the Refs [1], [5] –[7], taking into account the corrections given in this version, including that – since it has maximal t-IC – that the Planck mass particles can be micro black holes that do not "vapour" and so can be some candidates for a dark matter particles.
But here some puzzle for given model appears. As that was pointed out above in full particle's algorithm only a part is "informational" and define the particle. Under relativistic "dropping out" the moment should come when "ballast" will run out totally. E.g. for the proton one can to obtain some estimate of upper bound of "informational" length from cosmic rays spectrum – that is the $2\pi$- Compton length of a proton with the energy $\sim 10^{20}$ eV. It is possible that an attempt to add next energy portion for such a proton should lead to destruction of the particle. But the Planck mass particle has utmost minimal length in the rest state and so it can not move at all. For such a case QM says that the position of a particle becomes uncertain so these particles should be "smeared" in the Space.
But from another hand that is correct only for a specific frame system. In any other frame system which moves with a velocity $V$ in the first one – the particles should: (1) – to move with $V$, and (2) – so to have some location with uncertainty $\sim$ de Brougle wave length…

## 4. DISCUSSION AND CONCLUSION

There are till now two main philosophical concepts – Materialism and Idealism. The struggle between the concepts became nonsensical when I. Kant gave the proof that it is impossible to prove the existence / non-existence of God. Idealism lost it's arguments, when Materialism never had the ones – it is based on the laws that are obtained (and are tested by using, in fact the same) by using **necessary**, **but non – sufficient criterion** of the experiment results repetition, from which there can not be obtained even the answer on the question – "why do at all some laws exist?".

The informational concept is free from these backwards. **It is sufficient only once** (1) - "to detect in an experiment" some set of some elements; and (2) – "to detect" that on this set some language exists that include the notions "there is (an element)" and "there are no (elements)", as well as simple set of other logical rules, then at once an "experimenter" can build this information concept, including the expansion of the "experiment results" on the infinity set, developing on this way the set's theory; at that – *always remaining inside of the limits of the concept.*

But the Materialism /Idealism struggle gone now on a new stage. Materialism got the argument that the Nature laws can be "material" – since the "laws" are inherent to the Information; but Idealism got the argument that "in the Beginning was Word - some logical initial postulate" and so got the right on the question "Who did this Word say?"

Nevertheless, just the Information is utmost fundamental and "utmost infinite" set. In our World (Universe), as well as in possible other Worlds, under sequence of interactions between the elements at each step (an ISs interaction) next "instant" World picture is created; these instant pictures we call the "Matter", "the objective reality". Since the set Information principally can not be formalised, even it's particular realiztions, as, e.g., our World, are bifurcate. One of the bifurcations was, rather possibly, the appearance of "Consciousness".

Since any "finite" IS, including our World, is sub-sets in Information, these IS (as well, of course any IS from this IS) are always "open" system that uninterruptedly interacts with (full?) Information set so the Markov sequence of the "Matter pictures" is a random process. That becomes utmost clearly on microworld scale, when *the wave function is some mathematical representation of the information* - that always - even up to the Beginning - *existed and exists*, (as, e.g., a set of "Feynman paths") *about past and future,* of what we call "a particle in the spacetime".

Note, also, that it is true for any IS - any IS has it's "wave function" when any IS "is smeared" in spacetime in future. Though corresponding wave function depends on much lager number of parameters then for a particle, so to obtain this function for e.g., a human, seems rather difficult…

The informational concept presented here, besides that transforms long philosophical dispute of Materialism and Idealism concepts, yet now has rather great heuristic potential, at least in physics. Yet now one can conjecture a number of ideas that seemed worthwhile to elaborate. Some applications relating to understanding of some fundamental physics principles are given above, else - e.g. - the photon, rather probably, is an (integrated) combination of two components having spin ½; the Planck mass particles may be primordial micro black holes that can constitute the dark matter, etc. – see ([1], [5] –[7]).

And, notwithstanding the "wildness" of the set Information it is, nevertheless, some mathematical object which can be studied already by using existent instruments – the set theory, the theory of the language, the synergetics, etc.

… Now a rather popular line in physics development appeared – the development of "the Theory of Everything", when this theory should unite 4 forces that are considered in physics and are studied in corresponding physics branches. The attempts to create "ToE" became more intensive after successful development of the theory that "united" EM and weak forces.

But it is evident, that such a "ToE" can not be the theory of everything – the experimental physics isn't going to stop. E.g. – electric charge of the "fundamental particle" electron, as well as of some quarks, is obviously "non-fundamental", so in some future on an accelerator some "electron quarks" (or relating particles) rather possibly can be detected, what will require introducing into physics corresponding next force end development of the "Theory of next Everything", and so on. Such a process can be rather long, but in reality, as it follows from this information concept, the ToE will be eventually the theory of the set "Information".

The main advantage of the concept suggested above is that the Information concept is strictly logically grounded and is self-consistent. The main backward - because of we are always in the Information set and, as it follows from the main advantage, can not go out from this set – we can not to answer (at least we now) – what is besides of the set "Information", if It exists?
Though there is also some pleasant advantage - anybody knows all about total existing Information set – any point of a human brain contains the information: "in this brain point there are no…". So a little thing remains – to learn - how to read this information?